\documentclass[AMA,STIX2COL]{MRM}
\articletype{Research article}%

\usepackage{amsfonts}

\historydates{}
\makeatletter\let\printjnlcitation\relax\makeatother
\begin{document}

\title{High-Temporal-Resolution Motion Correction in Magnetic Resonance Fingerprinting Using a Quantitative Scout and Compact Spiral Navigators 
}

\author[1]{Aizada Nurdinova}{}
\author[1]{Xiaozhi Cao}{}
\author[2]{Daniel R. Abraham}{}
\author[3]{Daniel Polak}{}
\author[1]{Nan Wang}{}
\author[1,4]{Xuetong Zhou}{}
\author[2]{Yimeng Lin}{}
\author[1,2,4]{Brian A. Hargreaves}{}
\author[1,2]{Kawin Setsompop}{}
\authormark{Nurdinova \textsc{et al}}

\address[1]{\orgdiv{Department of Radiology}, \orgname{Stanford University}, \orgaddress{\city{Stanford}, \state{CA}, \country{USA}}}

\address[2]{\orgdiv{Department of Electrical Engineering}, \orgname{Stanford University}, \orgaddress{\city{Stanford}, \state{CA}, \country{USA}}}

\address[3]{\orgname{Siemens Medical Solutions USA, Inc.}, \orgaddress{\city{Malvern}, \state{PA}, \country{USA}}}

\address[4]{\orgdiv{Department of Bioengineering}, \orgname{Stanford University}, \orgaddress{\city{Stanford}, \state{CA}, \country{USA}}}

\presentaddress{Aizada Nurdinova, Department of Radiology Stanford University, Stanford, CA 94305, USA. 
\newline Email: nurdaiza@stanford.edu}

\finfo{\textbf{This work was partially funded by R01MH116173; R01EB019437; R01HD114719; R21EB038677.}}

\abstract[Abstract]{
\section{Purpose}
Motion correction in MRF helps preserve the accuracy of quantitative maps; however, existing approaches provide motion updates only every 7--8\,seconds. We propose a navigation framework integrating compact k-space navigators throughout the MRF acquisition, enabling sub-second motion estimation at minimal sequence overhead.

\section{Methods}
A 3D spiral-projection MRF sequence was augmented with three orthogonal spiral navigators inserted every 0.5\,seconds, enabling motion estimation by comparing navigator signals with Quantitative Scout (Q-scout) data, i.e., motion-free low-resolution k-space with matching contrast evolution. The Q-scout is obtained via a rapid calibration during the dummy preparation period, incurring no additional scan time. Motion estimation is formulated as dictionary matching in a discriminant subspace with optimization refinement. The method was evaluated in simulation and in vivo for 1\,mm isotropic brain 3D MRF at 3\,T.

\section{Results}
Across 35 motion-corrupted acquisitions with motion-free references available, the proposed motion correction reduced MRF reconstruction NRMSE by 7.1\% and increased SSIM by 0.085. Motion estimates aligned with 8\,second temporal-rate image-based navigation (mean absolute difference of 0.15\,mm and $0.23^\circ$), while the proposed method provided higher temporal resolution and improved motion correction.

\section{Conclusion}
The proposed framework enables robust motion navigation in MRF at 0.5\,second temporal resolution with minimal sequence overhead. By using contrast-consistent modeling and efficient inference, it improves the reliability of quantitative MRI under rapid, unpredictable motion.
}

\keywords{3D MR fingerprinting, retrospective motion correction, k-space navigators, contrast-matched navigation, motion-dictionary matching, discriminant analysis }

\maketitle

\section{Introduction}
Magnetic resonance fingerprinting (MRF) enables simultaneous estimation of tissue 
properties such as $T_1$, $T_2$, and proton density from a single acquisition with 
pseudorandomized sequence parameters, by matching voxel-wise signal evolutions to a 
simulated dictionary \cite{Ma2013_MRF_Nature}. Advances in acquisition design and 
efficient reconstruction have increased robustness while reducing scan times, 
supporting translation of MRF toward routine clinical deployment 
\cite{Gaur2023_MRF_ClinicalReview}.

Patient motion remains a major limitation in MRI, affecting approximately $20\%$ of 
examinations \cite{ANDRE2015689}. While MRF exhibits partial intrinsic robustness 
through frequent central k-space sampling and pattern-matching estimation 
\cite{Ma2013_MRF_Nature,Pierre2015_MRF_Multiscale}, motion can perturb temporal 
signal evolution and bias quantitative estimates even without visible image artifacts 
\cite{Gaur2023_MRF_ClinicalReview,Callaghan2015_PMC_QMRI,Noeth2014_T2starMotionCorrection,
Yu2018_MRI_MRF_MotionSensitivity}.

In brain imaging, motion can be approximated as rigid with six degrees of freedom 
(DOF), motivating a wide range of motion correction (MoCo) strategies. External 
tracking using optical systems \cite{zaitsev2015optical,qin_optical}, pilot tone 
\cite{ludwig2021pilot,brackenier_pt}, beat pilot tone \cite{anand_bpt}, and NMR 
markers \cite{sengupta_nmr,Eschelbach_nmr,haeberlin_nmr,Aranovitch_nmr} provide 
accurate estimates but require additional hardware. Motion can alternatively be 
estimated directly from MR signals. Early model-based approaches formulated MoCo as 
joint or alternating estimation during reconstruction 
\cite{Haskell2018TAMER,namer2019,CorderoGrande20183DMoCo}, but these are non-convex 
and computationally demanding. To improve efficiency, the SAMER and related frameworks 
introduced a motion-free scout image as a reference to guide motion optimization 
directly from k-space data \cite{Polak2021SAMER,fid_motion,fid_shim}.

K-space navigators, compact acquisitions dedicated to motion sensing and interleaved 
per sequence unit without disrupting the main acquisition, enable frequent motion 
updates. A variety of trajectories have been proposed, including cloverleaf 
\cite{vanderKouwe2006Cloverleaf}, butterfly \cite{butterfly2012}, FID 
\cite{fid_motion,fid_shim}, spherical \cite{Hewlett2024SNAV}, orbital 
\cite{Ulrich2023ServoNav,serger2026motion,onav_ward}, and SPINS 
\cite{Brackenier2023QUEEN}. These methods typically rely on a short calibration 
acquisition or simulation to correlate newly acquired navigator signals for motion estimation. SAMER 
was also extended to recover motion from 2--4 interleaved k-space line navigators per 
echo train \cite{Polak2022MotionGuidance}.

Existing MRF-specific MoCo strategies mostly rely on volumetric navigators. 
Sliding-window reconstruction-based registration, proposed for both 2D and 3D MRF, 
achieves motion updates every acquisition group ($\sim$7\,s) 
\cite{Xu2019_MRF_RigidMotion_SlidingWindow,Kurzawski2020_RetrospectiveMRF_Motion}. 
Alternatively, low-resolution image navigators~\cite{Cao2022} or fat 
navigators~\cite{Hu_Fatigator,Chen2020_3DMRF_FatNavigator} acquired at the end of each MRF 
acquisition group have been proposed, the latter building on the FatNav 
concept~\cite{Gallichan_fatnav}, where fat-selective excitation enables 
high-resolution motion estimation. These approaches ensure consistent contrast 
across registered volumes without disturbing the signal evolution, reaching 
motion sensitivities of $\sim$0.05\,pixel/$^\circ$, but remain limited to 
7--8\,s temporal resolution.

Extending MRF MoCo to higher temporal resolution by inserting k-space navigators 
throughout the MRF acquisition groups introduces a key challenge: the rapidly varying 
contrast of the MRF sequence produces navigators with time-varying signal that cannot 
be correlated against a single static reference, as contrast- and motion-induced 
signal changes are otherwise indistinguishable. The 
Quantitatively Enhanced Navigation (QUEEN) framework addresses this and builds on the SAMER approach by using 
a dedicated quantitative scout (Q-scout) scan to obtain low-resolution tissue 
parameter maps, which combined with analytical signal expressions enable generation of 
contrast-matched scout images 
\cite{Brackenier2023QUEEN}. Although QUEEN has demonstrated 
accurate motion tracking in GRE and MPRAGE sequences, its application to MRF remains 
unexplored.

We propose a sub-second temporal-resolution motion navigation framework for 3D spiral 
projection MRF (\textbf{3D SPI MRF}). A Q-scout is acquired during the sequence dummy 
preparation period without added scan time. Three orthogonal spiral navigators are 
interleaved every 0.5\,s during acquisition, enabling rapid pose updates at 8\% scan 
efficiency cost. Motion is estimated using motion-dictionary matching for global initialization, bypassing local minima, followed 
by lightweight optimization in a discriminant subspace designed to emphasize 
motion-induced variations while reducing sensitivity to model imperfections. 
Simulations demonstrate accurate motion estimation with substantial computational 
speedups over conventional optimization. In vivo validation across 35 
motion-corrupted volunteer scans shows consistent artifact reduction with motion 
trajectories agreeing with lower-temporal-resolution image-navigator approaches.

\section{Methods}
We first provide background on the underlying 3D SPI MRF acquisition and reconstruction, then describe the proposed motion navigation framework, comprising a Q-scout, compact navigators, and motion estimation approach, and finally the implementation and evaluation setup. While demonstrated on 3D SPI MRF, the framework applies to other MRF variants and MRI sequences.

\subsection{3D SPI MRF}
\begin{figure*}
\centerline{\includegraphics[width=\textwidth]{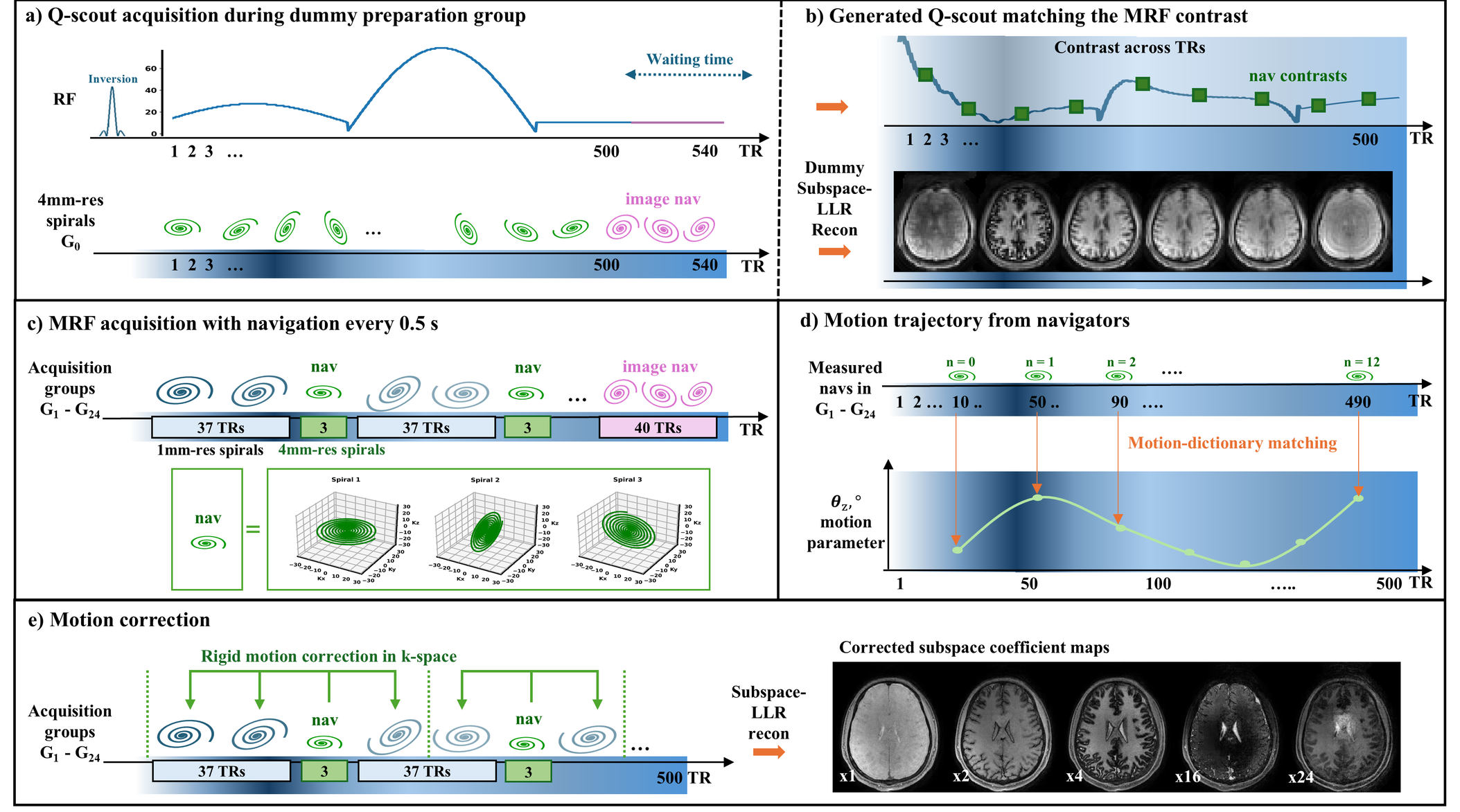}}
\caption{Proposed motion navigation pipeline for 3D spiral-projection MRF. a) Q-scout is acquired during the dummy scan without added scan time using a 4-mm isotropic resolution SPI-MRF scan with the same variable RF flip-angle train as in the main MRF acquisition. b) The Q-scout forms a low-resolution, temporally resolved image volume with contrast evolution across 500 TRs matched to the main MRF acquisition (shown as blue color gradient). c) During the MRF acquisition, three orthogonal spiral navigators (trajectories shown) are acquired every 40 TRs to provide motion estimation at a temporal resolution of 0.5\,s. d) Motion estimation: navigator signals are matched against a Q-scout-derived motion dictionary in a discriminant subspace, followed by lightweight optimization to recover the rigid motion trajectory. e) The estimated motion parameters are applied to correct the corresponding k-space data, followed by subspace-LLR reconstruction to obtain motion-corrected subspace coefficient maps, which are subsequently fitted to generate tissue parameter maps.\label{fig:sequence}}
\end{figure*}

Figure~\ref{fig:sequence}a,c shows the sequence diagram for 3D SPI MRF, which enables rapid whole brain $T_1$, $T_2$, and proton density (PD) mapping at 1~mm isotropic resolution in 2-3 minutes. The acquisition follows an inversion recovery pulse with a variable flip-angle train, generating controlled contrast modulation while sampling highly efficient non-Cartesian 3D spiral rotations that yield incoherent space-time undersampling. Each acquisition group contains 500 TRs with a duration of 8 seconds and is repeated with different spiral projection samplings to ensure sufficient k-space coverage at each TR. For instance, 1~mm isotropic brain imaging requires 16 acquisition groups to achieve adequate sampling per contrast timepoint~\cite{Cao2022}.

Reconstruction (Figure~\ref{fig:sequence}e) is performed using temporal subspace modeling~\cite{zhipei2007} across the 500 TRs. Five temporal subspace bases ($N_{ {coef}} = 5 $) are sufficient to represent the contrast evolution and enable high-quality parameter mapping~\cite{Cao2022}. Using a precomputed contrast subspace bases $\boldsymbol{\Phi}$, temporally resolved images across TRs are expressed from coefficient maps $\boldsymbol{q}$ as:
\begin{equation}
\boldsymbol{x} = \boldsymbol{\Phi}\,\boldsymbol{q},
\label{eq:contrast_subspace} 
\end{equation}
where $\boldsymbol{x} \in \mathbb{C}^{N_{ {contrast}} \times N_{ {pixels}}}$ with $N_{ {contrast}}=500$, $\boldsymbol{q} \in \mathbb{C}^{N_{ {coef}} \times N_{ {pixels}}}$, and $\boldsymbol{\Phi} \in \mathbb{C}^{N_{ {contrast}} \times N_{ {coef}}}$. The coefficient maps $\mathbf{q}=\{q_1,\ldots,q_5\}$ are reconstructed by solving
\begin{equation}
\hat{\mathbf{q}}
=
\arg\min_{\mathbf{q}}
\left\|
\mathbf{y}
-
 {(P\,F)\,C\,\Phi}\,\mathbf{q}
\right\|_2^2
+
\lambda_{\text{LLR}}
\sum_{b \in \Omega}
\left\|
\mathcal{R}_b(\mathbf{q})
\right\|_* ,
\label{eq:subspace_llr}
\end{equation}
where $\mathbf{y}$ is the measured multi-coil non-Cartesian k-space data, $(P\,F)$ denotes the non-uniform fast Fourier transform (nuFFT), and $C$ represents coil sensitivity maps. The second term enforces locally low-rank (LLR) regularization, where the $\mathcal{R}_b$ operator extracts a fixed-size patch $b$, $\Omega$ is the set of patches, and $\|\cdot\|_*$ is the nuclear norm.

Low-resolution \textbf{image navigators} are incorporated in 3D SPI MRF by acquiring 
40 TRs at a constant flip angle of 10° over 0.5\,s at the end of each acquisition 
group, before the 1.2\,s magnetization recovery period (Figure~\ref{fig:sequence}a). 
The resulting 4\,mm resolution images are registered across groups to estimate motion. While robust, this strategy limits updates to 8\,s temporal resolution.

\subsection{Navigation Acquisition Design}

In designing the navigation framework, we follow prior SAMER work, which demonstrates 
that a scout with 4$\times$ lower resolution than the imaging sequence combined with compact 
navigators acquiring 2--4 k-space lines within the scout support region enables robust 
motion estimation \cite{Polak2021SAMER}.

\subsubsection{Q-scout}

A quantitative scout for motion estimation requires low-resolution tissue parameter 
maps that enable generation of contrast-matched scout images at navigation TRs within MRF. This Q-scout can be obtained using a low-resolution 3D SPI 
MRF acquisition, where a single acquisition group in 8\,s suffices to reconstruct 
good quality 4\,mm isotropic volumes (Figures~\ref{fig:sequence}b 
and~\ref{fig:q_scout}). Standard 3D SPI MRF includes one dummy preparation group 
before acquisition to reach steady state, so a Q-scout with signal evolution matched 
to the main acquisition would naturally be placed as the second group, resulting in 
8\,s of added calibration time. To avoid adding scan time, we instead synthesize the Q-scout from the dummy preparation group.

Since the dummy group begins from fully recovered magnetization rather than steady state, 
a dedicated temporal subspace $\Phi_{\text{dummy}}$ is computed to account for the 
differing signal evolution during reconstruction. The resulting subspace coefficient 
maps are dictionary-matched to estimate tissue parameter maps ($T_1$, $T_2$, PD), 
from which Q-scout subspace coefficient maps with contrast evolution matched to the main 
acquisition subspace $\Phi$ are synthesized. 

\subsubsection{Spiral Navigators}

Navigators use the same 4\,mm spiral trajectories as the Q-scout acquisition, ensuring 
both share similar imperfections from gradient errors, static $B_0$ inhomogeneity, 
eddy currents, and $T_2^*$ modulation. Three orthogonal spiral trajectories are 
selected from the Q-scout readouts to achieve sensitivity to all 6 rigid-body motion 
parameters (Figure~\ref{fig:sequence}c). Flip angles match those of the underlying 
MRF sequence at each navigator TR to preserve the signal evolution.

Spiral navigators are interleaved every 40 TRs starting at TR index 10 within each 
high-resolution MRF acquisition group, yielding 13 navigation timepoints per 500-TR 
group and an update rate of 0.5\,s. The three orthogonal spiral readouts per 
timepoint result in a scan efficiency reduction of $3\times13/500\approx8\%$.

\subsection{Motion Estimation}

With the sequence modifications enabling sub-second navigation using a contrast-consistent Q-scout, we now describe the motion estimation. We first formulate it via optimization following prior work \cite{Brackenier2023QUEEN,Polak2021SAMER,CorderoGrande20183DMoCo,Haskell2018TAMER}, then propose to reformulate it using dictionary matching to improve computational efficiency.

Using Eq.~\ref{eq:contrast_subspace} and given the Q-scout $\boldsymbol{q_{ {scout}}}$ and navigator measurements $\mathbf{s}_{(g,n)}$ acquired at acquisition group $g$ and navigation timepoint index $n$, motion estimation is formulated as:

\begin{equation}
\begin{aligned}
\hat{\boldsymbol{\theta}}_{(g,n)}
={}&
\arg\min_{\boldsymbol{\theta}_{(g,n)}}
\Bigl\|
\mathbf{s}_{(g,n)}
-{}
\\
&\quad -
({P}_{nav}\,F)\,C\,T_{\boldsymbol{\theta}_{(g,n)}}
\left(
\boldsymbol{\Phi}_n \mathbf{q_{scout}}
\right)
\Bigr\|_2^2
+
\lambda_{\theta} R(\boldsymbol{\theta}),
\end{aligned}
\label{eq:motion_optim}
\end{equation}

where $T_{\boldsymbol{\theta}}$ denotes the rigid-motion operator, $C$ are coil sensitivity maps, and (${P}_{\text{nav}}\,F$) is the nuFFT operator with the navigator trajectory. The regularizer $R(\boldsymbol{\theta})$ enforces temporal smoothness on the estimated trajectory across navigator timepoints, making the problem non-separable across $g$ and $n$.

Although prior work shows that the problem in Eq.~\ref{eq:motion_optim} is locally convex within motion ranges relevant for MRI (approximately $\pm 20$~mm and $\pm 20^\circ$)~\cite{Polak2021SAMER}, it can still be poorly conditioned, leading to slow convergence. This issue is exacerbated when only three navigator readouts are available per pose, as limited sampling increases sensitivity to noise and model inaccuracies, and may introduce coupling between motion degrees of freedom (Figure S1). 

To address these challenges, we formulate motion estimation as a dictionary matching problem over discretized motion states. This enables an efficient global search that avoids reliance on local loss curvature, thereby mitigating the effects of poor conditioning and reducing sensitivity to initialization (see Figure S1a). A subsequent lightweight optimization step refines the estimate to correct for discretization error.

\begin{figure*}
\centerline{\includegraphics[width=\textwidth]{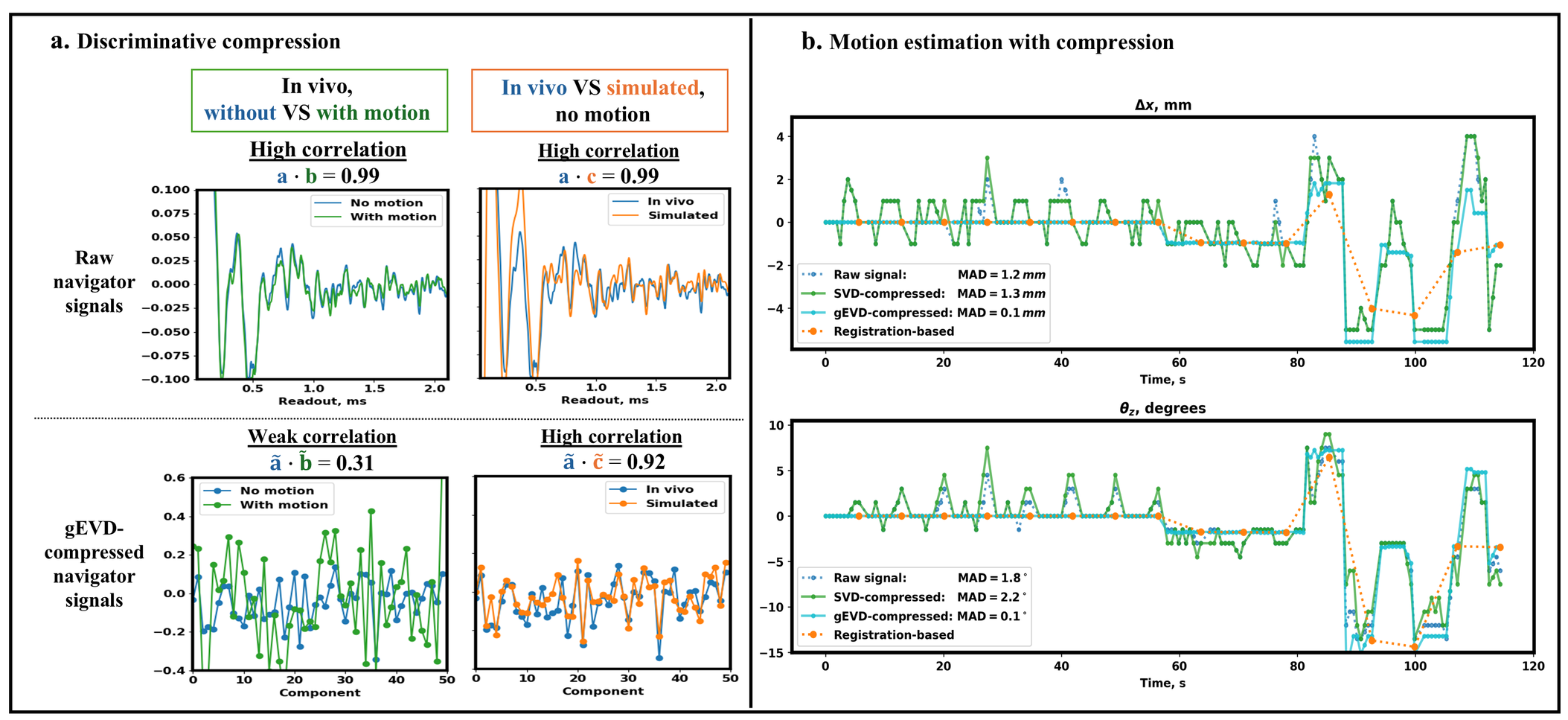}}
\caption{Validation of the proposed gEVD-based discriminant subspace for robust in vivo motion estimation. a) Without compression, motion-free simulated and in vivo navigator signals are highly correlated ($\sim$0.99), as are in vivo signals with and without motion, making motion and modeling errors hardly distinguishable. A gEVD basis constructed from the simulated motion dictionary and Q-scout no-motion navigators enhances separability between motion states while preserving simulation-to-in-vivo agreement. After compression, correlation between in vivo signals with and without motion drops to 0.32, while no-motion simulated and in vivo signals remain well matched (0.92). b) Motion estimation in a slow-motion experiment (subject was instructed to remain still for the first minute). Raw and SVD-compressed motion dictionary matching (DM) show variability during no-motion periods. gEVD-compressed signals produce stable estimates consistent with lower-temporal rate image registration estimates, while accelerating dictionary matching by $\sim400\times$. Motion-corrected reconstructions are shown in Figure S3.
\label{fig:compress}}
\end{figure*}

\subsubsection{Dictionary-Based Reformulation}

The motion dictionary over a discrete motion grid $\boldsymbol{\Theta}$ can be computed on subspace coefficient maps\cite{tamir_t2shuffle} and projected to each navigation timepoint index $n$ following Eq.~\ref{eq:contrast_subspace}:

\begin{equation}
D_n(\boldsymbol{\Theta})
=
\boldsymbol{\Phi}_n
\left(
\left\{
 {P_{nav}\,F\,C\,T}_{\boldsymbol{\theta}}\,\mathbf{q_{scout}}
\right\}_{\boldsymbol{\theta} \in \boldsymbol{\Theta}}
\right).
\end{equation}

with dimensions $N_{\text{motion}} \times (N_{\text{ch}} \cdot N_{\text{t}} )$, where $N_{\text{ch}}$ is the number of coil channels, and $N_{\text{t}}\sim3\times 1600$ is the total number of readout timepoints using three concatenated orthogonal spirals.

Motion parameters are estimated by matching navigator signals $s_{(g, n)}$ to the corresponding contrast-specific dictionary:
\begin{equation}
\hat{\boldsymbol{\theta}}_{(g,n)}^{ {DM}}
=
\arg\max_{\boldsymbol{\theta}\in\boldsymbol{\Theta}}
\left\langle D_n(\boldsymbol{\theta}), \mathbf{s}_{(g,n)} \right\rangle,
\end{equation}
using the normalized complex similarity metric
\begin{equation}
\left\langle \boldsymbol{d}_n, \boldsymbol{s}_{(g,n)} \right\rangle
=
\frac{
|\boldsymbol{d}_n^{H}\boldsymbol{s}_{(g,n)}|
}{
\|\boldsymbol{d}_n\|_2\,\|\boldsymbol{s}_{(g,n)}\|_2
}.
\label{eq:dot_metric}
\end{equation}
Where $(g,n) \in \{0,\ldots,N_{\text{groups}}-1\} \times \{0,\ldots,N_{\text{navi}}-1\}$, with $N_{\text{groups}}$ denoting the number of acquisition groups in the SPI MRF acquisition and $N_{\text{navi}}$ the number of navigator contrast points per group.

Simulations of the motion dictionary reveal a pronounced low-rank structure (Figure S2a), enabling effective dimensionality reduction, which can reduce memory requirements and significantly accelerate motion estimation.

\subsubsection{Discriminant Extraction}
Global mismatch between measured and simulated navigator signals arising from 
unmodeled effects, such as contrast subspace approximation error, eddy 
currents~\cite{serger2026motion}, or $B_0$ inhomogeneity, can produce signal 
variations comparable to those induced by motion (Figure~\ref{fig:compress}a), 
degrading motion state discriminability. To address this, we introduce a discriminative 
compression of navigator signals based on linear discriminant analysis (LDA) via 
generalized eigenvalue decomposition (gEVD)~\cite{bishop2006pattern,fukunaga1990introduction}, 
which separates motion-induced variability from unmodeled nuisance effects. 

The discriminant basis is constructed independently at each navigator contrast timepoint 
to account for contrast-dependent model mismatch. To reduce computational cost during 
basis construction, compression is performed along the temporal dimension only, 
though the coil dimension could alternatively be explicitly retained and jointly 
compressed within the same framework.

The discriminants are computed from simulated dictionary signals across motion states 
and experimentally acquired no-motion navigators from the Q-scout. To enable the 
latter, the Q-scout acquisition is modified such that at each navigator timepoint 
within the dummy group (every 40 TRs starting at TR index 10), the same three orthogonal 
spiral readouts are acquired.

At a fixed contrast, the motion dictionary $D$ of shape $N_{\text{motion}} \times N_{\text{ch}} \times 
N_{\text{t}}$ contains rows $d_m$ of size $N_{\text{ch}} \times N_{\text{t}}$ 
corresponding to each motion state $m$ with Gaussian readout noise with measured 
covariance $\Sigma$ (can be estimated from a noise scan), and $y_0$ of size $N_{\text{ch}} \times N_{\text{t}}$ denotes 
the in vivo no-motion measurement. Mean signals per state and global mean are defined as:
\begin{equation}
\begin{aligned}
    \mu^0 &= \frac{y_0 + d_0}{2}, \\
    \mu^m &= d_m, \qquad m = 1, \dots, M-1,
\end{aligned}
\end{equation}
\begin{equation}
    \mu = \frac{1}{M}\!\left(\sum_{m=1}^{M-1} \mu_m + \mu^0\right),
\end{equation}
where the zero-motion state $m=0$ is treated as a combination of simulated ($d_0$) 
and measured ($y_0$) signals, explicitly modeling simulation-measurement mismatch for 
intra-state variability. The between-state $S_B$ and within-state $S_W$ scatter 
matrices capture motion-related versus noise- and model-imperfection-induced signal 
variation, and are constructed as:
\begin{equation}
\begin{aligned}
    S_B &= \sum_{m=0}^{M-1}\,\sum_{c=0}^{C-1}(d_{m,c} - \mu_c)(d_{m,c} - \mu_c)^H, \\
    S_W &= \sum_{m=0}^{M-1}\,\sum_{c=0}^{C-1}n_c\,n_c^H + S_W^0 
         = M\operatorname{tr}(\Sigma)I_T + S_W^0, \\
    S_W^0 &= \sum_{c=0}^{C-1}\frac{1}{2}(y_{0,c} - d_{0,c})(y_{0,c} - d_{0,c})^H,
\end{aligned}
\end{equation}
where $n_c \sim \mathcal{N}(0, \Sigma_{c,c})$ is the Gaussian readout noise per coil, 
$S_B, S_W$ have dimensions $N_{\text{t}} \times N_{\text{t}}$ and capture 
variance along the readout dimension. The term $M\operatorname{tr}(\Sigma)I_T$ ensures $S_W$ is full rank and stabilizes its inversion in gEVD.

We seek a linear compression matrix $U$ of shape $K \times N_{\text{t}}$, 
$K < N_{\text{t}}$, maximizing motion separability relative to nuisance variability:
\begin{equation}
    \hat{U} = \arg\max_U\operatorname{tr}\!\left[(US_WU^H)^{-1}(US_BU^H)\right],
\end{equation}
whose rows are given by the $K$ leading eigenvectors of the generalized eigenvalue 
problem
\begin{equation}
    S_B\,u_{i}^T = \lambda_i\,S_W\,u_{i}^T,
\label{eq:gevd}
\end{equation}
solvable directly or via whitening of $S_W$ followed by SVD~\cite{bishop2006pattern}.

While LDA is Bayes-optimal under Gaussian noise alone, the gEVD still identifies 
the best linear subspace separating motion states relative to the total nuisance 
variability in $S_W$, although the resulting boundary is not guaranteed to be optimal 
due to the unknown structure of $S_W^0$. Empirically, this compression is shown 
to reduce estimation bias in in vivo data (Figure~\ref{fig:compress}).

\begin{figure*}
\centerline{%
\includegraphics[width=0.65\textwidth]{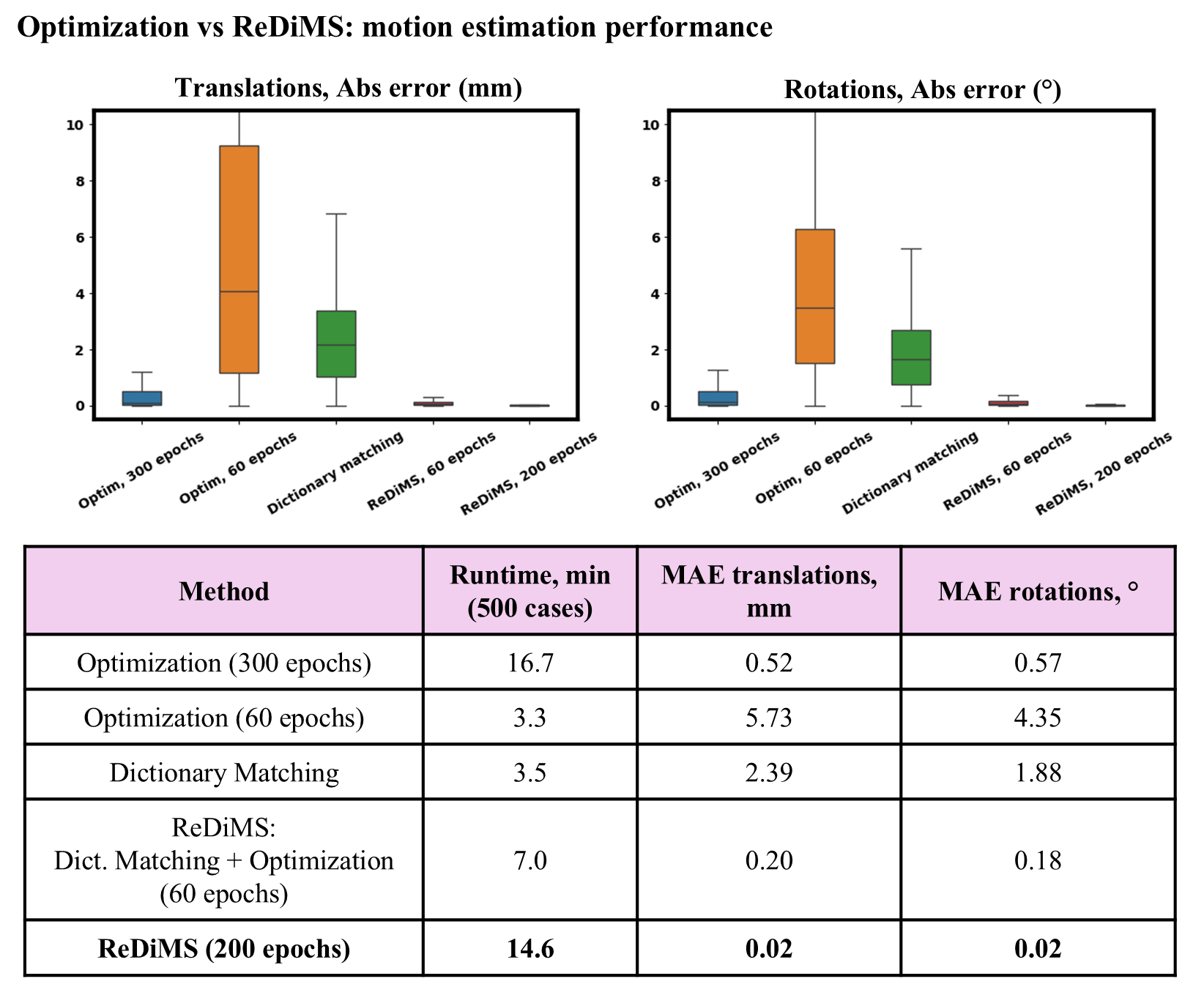}%
\hspace{1em}%
\parbox[b]{0.3\textwidth}{%
\caption{ReDiMS performance compared with conventional optimization in simulations. 500 navigator signals were simulated with random 6-DOF motion within $\pm$12.5\,mm$/^\circ$ and added Gaussian noise, and five gEVD-compressed motion estimation approaches were compared. 60-epoch optimization from zero initialization failed to converge in 3.3\,min. 300-epoch optimization reached MAE of $\sim$0.5\,mm$/^\circ$ in 16.7\,min. Dictionary matching alone achieved MAE of 2.4\,mm and 1.9$^\circ$ in 3.5\,min, with accuracy limited by the motion-grid spacing. ReDiMS with 60-epoch optimization refinement achieved MAE of $\sim$0.2\,mm$/^\circ$ in 7.0\,min, and 200-epoch refinement further reduced MAE to 0.02\,mm$/^\circ$ in 14\,min.\label{fig:sims}}%
}}
\end{figure*}

\subsubsection{ReDiMS Framework}

Proposed motion estimation is performed via coarse dictionary matching followed by continuous optimization refinement, both in the discriminant subspace for robustness to modeling mismatch. We refer to this approach as \textbf{ReDiMS} (\textbf{Re}fined \textbf{Di}ctionary \textbf{M}atching in a discriminant \textbf{S}ubspace).

For each navigator contrast timepoint $n$, a compressed motion dictionary is generated over a wide motion range using a coarse parameter grid:
\begin{equation}
{\tilde{D}_n}(\boldsymbol{\Theta})
=
{U_n}\,{D_n}(\boldsymbol{\Theta}),
\end{equation}

Navigator signals acquired during the MRF acquisition are projected onto the same contrast-specific basis, resulting in signal representation of length $K$:
\begin{equation}
\tilde{\mathbf{s}}_{(g,n)}
=
{U_n}\, \mathbf{s}_{(g,n)},
\label{eq:project}
\end{equation}

Motion parameters at each navigator timepoint are first estimated via regularized discrete dictionary fitting:
\begin{equation}
\begin{aligned}
{\boldsymbol{\hat\theta}}_{(g,n)}^{ {DM}}
=
\arg\max_{{\boldsymbol{\theta}} \in \boldsymbol{\Theta}}
&\left\langle
\tilde{D}_n({\boldsymbol{\theta}}),
\tilde{\mathbf{s}}_{(g,n)}
\right\rangle
+{}
\\
&+
\lambda_{DM}
\left\|
{\boldsymbol{\theta}}_{(g,n)}
-
{\boldsymbol{\theta}}_{(g,n-1)}
\right\|_2^2,
\end{aligned}
\label{eq:dot}
\end{equation}
where the inner product measures similarity in the compressed space and temporal 
regularization prevents large jumps in the motion trajectory between neighboring 
timepoints.

As the discrete estimates are limited by the resolution of the motion grid, they are subsequently refined via continuous optimization in discriminant subspace, initialized at ${\boldsymbol{\theta}}^{ {DM}}$:
\begin{equation}
\begin{aligned}
\boldsymbol{\hat\theta}
={}&
\arg\min_{\boldsymbol{\theta}} \;
\sum_{(g,n)}
\Bigl\|
U_n
\Bigl(
\mathbf{s}_{(g,n)}
-{}
\\
&\quad -
{(P_{nav}\,F)\,C\,T}_{\boldsymbol{\theta}_{(g,n)}}
\,\boldsymbol{\Phi_n}\,\boldsymbol{q_{ {scout}}}
\Bigr)
\Bigr\|_2^2
+
\lambda_1
\left\|
\nabla \boldsymbol{\theta}
\right\|_2^2,
\\
&\quad
\text{initialized with } \boldsymbol{\theta}_{ {init}} = \boldsymbol{\hat\theta}^{ {DM}}
\end{aligned}
\label{eq:refine}
\end{equation}

\subsection{Experimental Setup}

\subsubsection{Compute Resources}

All experiments were run on a dual-socket workstation with Intel Xeon Gold 5320 CPUs and an NVIDIA RTX A6000 GPU (48\,GiB).

\subsubsection{Acquisition}

The 3D SPI MRF sequence with image navigators at the end of each acquisition group, Q-scout acquisition and spiral motion navigators was implemented using the KS Foundation framework~\cite{ksfoundation} and deployed on a 3T GE Signa Premier system (GE Healthcare, Madison, WI) using a 32-channel head receiver coil.

Ten healthy volunteers were scanned under institutional review board (IRB) approval with written informed consent. For each subject, one motion-free reference acquisition and 3-4 acquisitions with deliberate motion were performed.

Prior work used 16 acquisition groups (2.5-minute scan time) for high-quality 1 mm isotropic brain tissue parameter mapping~\cite{Cao2022}. Here, we extend this to 24 groups (3-minute scan time) in motion-corrupted settings to improve reconstruction conditioning for large-motion retrospective correction when estimates are accurate.

To evaluate the generated Q-scout quality, a second dummy group with contrast evolution identical to the main acquisition was acquired. This additional acquisition was used solely for evaluation and is not included in a deployed protocol.

\begin{figure*}
\centerline{%
\includegraphics[width=0.65\textwidth]{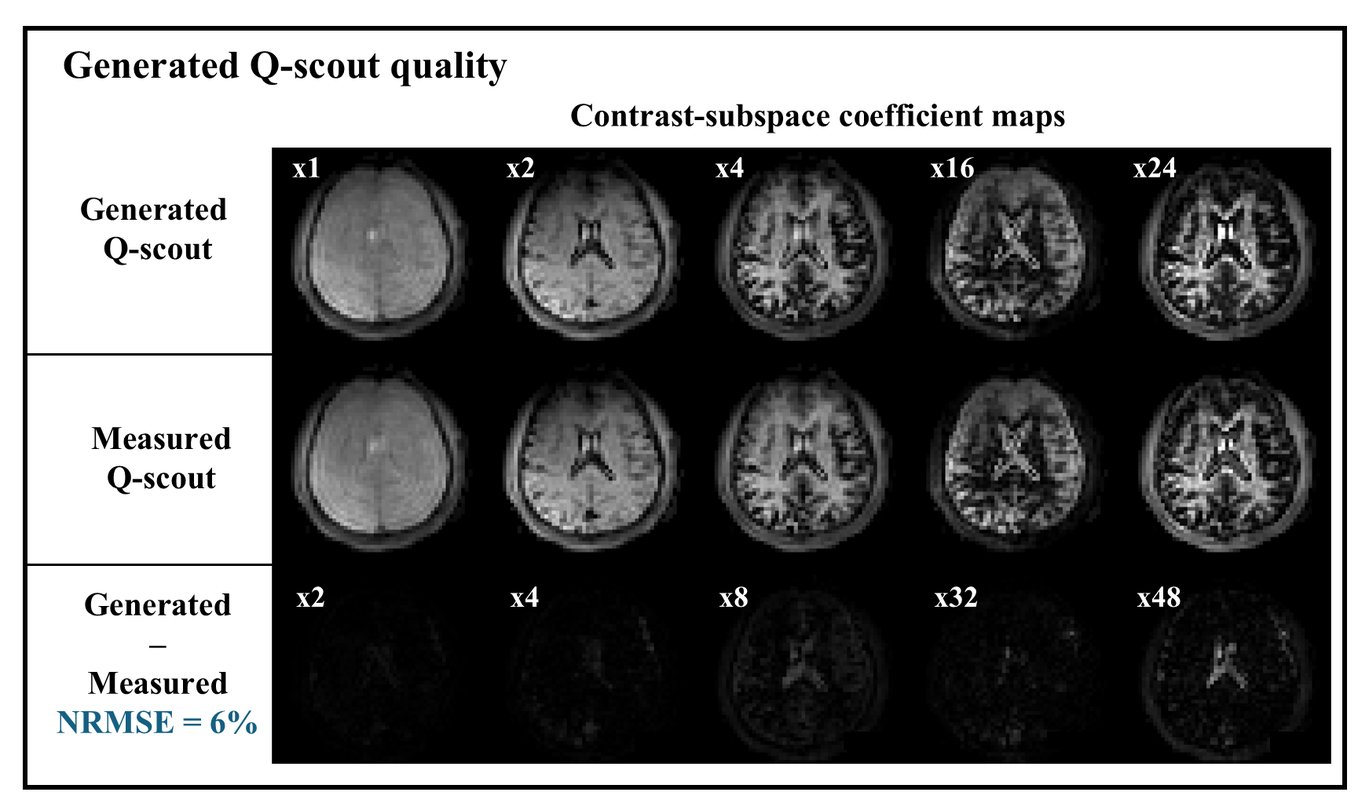}%
\hspace{1em}%
\parbox[b]{0.3\textwidth}{%
\caption{Validation of Q-scout generation from a dummy scan with different initial $M_z$ state than the main MRF acquisition. The Q-scout is generated by reconstructing tissue parameter maps from the dummy acquisition and mapping them to subspace coefficients in the main acquisition's temporal subspace. Generated Q-scout subspace coefficients are compared with ground truth coefficients from a second dummy scan (acquired for this validation only). Strong agreement is observed, with differences (x2 scale) primarily in CSF and skull regions for later coefficients with reduced signal contributions.\label{fig:q_scout}}
}}
\end{figure*}

\subsubsection{Reconstruction}
The MRF tissue parameter dictionary was precomputed using the extended phase graph (EPG) formalism~\cite{weigel2015} and compressed to five SVD components~\cite{Cao2022}. Reconstruction followed Eq.~\ref{eq:subspace_llr}, with coil sensitivity maps estimated using ESPIRiT~\cite{Uecker2014} and locally low-rank regularization applied with $10\times10$ patches. To accelerate reconstruction, five SVD-compressed coil sensitivity maps capturing 96\% of the signal variance were used. Implementation was performed in PyTorch~\cite{paszke2019arxiv} with support from SigPy~\cite{ong2019sigpy}. The regularization parameter $\lambda_{\text{LLR}}$ was tuned empirically per subject for static and motion-corrupted/corrected reconstructions.

Rigid {motion correction} was performed directly in k-space by modulating the data with a phase ramp and rotating the sampling coordinates~\cite{bammer2007augmented, cruz2019rigid, Xu2019_MRF_RigidMotion_SlidingWindow, godenschweger2016motion}. The estimated motion trajectory was applied in a piecewise-constant manner, with each estimate applied at the acquisition interval midpoint.

{Tissue parameter maps} were obtained by voxel-wise template matching of the reconstructed subspace coefficient maps to the MRF tissue parameter dictionary~\cite{Cao2022}. To account for transmit field inhomogeneity, $B_1$ correction was incorporated using a $B_1$-map estimated directly from the MRF data with the SAFE method~\cite{gao2023safe}.

For Q-scout generation, a dummy-group tissue parameter dictionary was simulated and used for subspace-LLR reconstruction, followed by parameter mapping. The resulting maps were reprojected into the main acquisition subspace to generate the Q-scout with contrast matched to navigator time points. Although this produces 500 contrast time points, only those corresponding to navigator acquisition times were retained for motion estimation.

Conventional image navigators in 3D SPI MRF~\cite{Cao2022} were used as a \textit{low-temporal-resolution reference}. Low-resolution spiral readouts (40~TRs) acquired at the end of each group were reconstructed using conjugate-gradient SENSE (CG-SENSE) with total variation regularization~\cite{Pruessmann1999}, yielding one navigator per group (8~s temporal resolution). Rigid registration using AFNI \texttt{3dvolreg}~\cite{cox1996afni} provided motion estimates. The final high-temporal-rate spiral navigators within each MRF group were acquired 10 TRs (125\,ms) prior to the image navigator, allowing comparison of estimates at these time points for validation.

\begin{figure*}
\centerline{\includegraphics[width=0.9\textwidth]{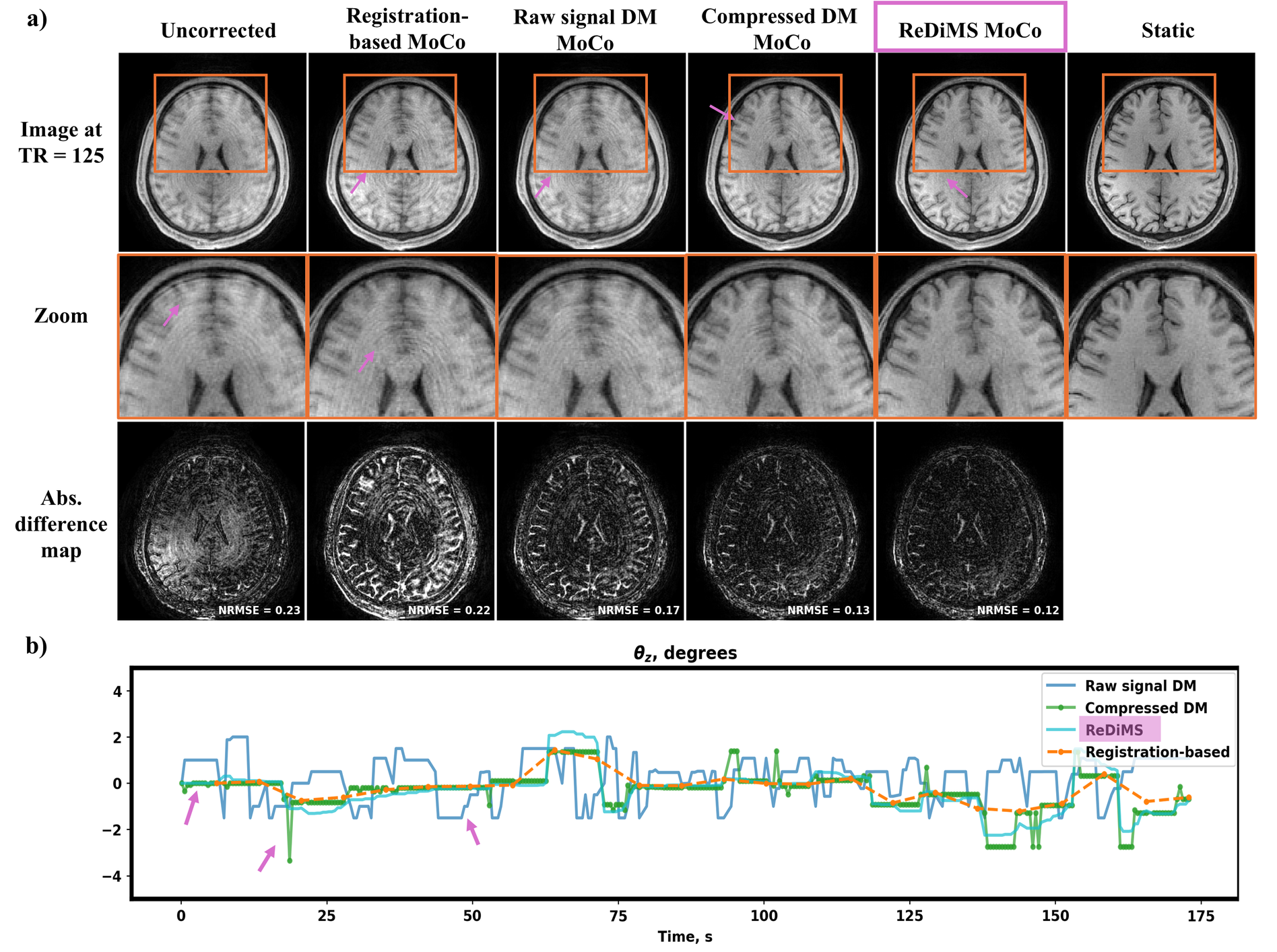}}
\caption{Comparison of motion estimation and correction on a representative in vivo motion-corrupted MRF using four different approaches: image registration-based, raw navigator-signal motion-dictionary matching (Raw signal DM), gEVD-compressed-signal motion-dictionary matching (Compressed DM), and compressed-signal motion-dictionary matching with continuous optimization refinement (ReDiMS). a) Time-resolved images at TR = 125 obtained from motion-corrected subspace reconstruction using the four approaches, with uncorrected and static references. Progressive artifact reduction and decreasing NRMSE are observed from registration-based to higher-temporal rate navigator-based correction, with further improvements from signal compression and refinement steps. b) Estimated motion trajectories are shown for $\theta_z$ parameters, where image registration results at low temporal resolution are assumed to be accurate anchor points for comparison. Pink arrows indicate where raw signal DM deviates from the reference, while gEVD compression aligns better with the reference, although some spikes are still noticeable. Refinement in ReDiMS further reduces discretization errors and regularizes the trajectories, aligning better with the reference method.\label{fig:ablation}}
\end{figure*}

\subsubsection{ReDiMS Implementation}

Compression and discriminant basis design were informed by simulation experiments in Figure S2. The discriminant basis $U$ was computed from a small dictionary with 3 grid steps per DOF, $\Theta_{\text{small}} = \{\theta \mid \theta_i \in \{-\Delta\theta_i, 0, \Delta\theta_i\},\ 0 \leq i \leq 5\}$, with $\Delta\theta = 0.5$\,mm/$^\circ$, as shown in Figure S2d.

The large matching dictionary had a fixed size of $|\Theta| = 5^6 \approx 15{,}000$ motion states, corresponding to 5 grid steps per DOF, and was defined on a uniform grid $\Theta$ with adaptive grid spacing $\Delta\theta \sim 0.5$--$6.0$\,mm/$^\circ$, selected based on the motion range estimated from image navigators.

Motion estimation proceeds as follows. Given $\mathbf{q_{\text{scout}}}$, 
SVD-compressed coil sensitivity maps $C$, navigator trajectories $P_{\text{nav}}$, 
measured navigator signals $\mathbf{s}_{(g,n)}$, no-motion navigator signals $y_0$, 
temporal subspace $\boldsymbol{\Phi}$, and noise covariance $\Sigma$:

\begin{enumerate}[leftmargin=16pt,label=(\arabic*)]
    \item Compute discriminant bases $U_n$ per contrast using $y_0$, $\Sigma$, 
    and dictionary on $\Theta_{\text{small}}$.
    \item Generate a large-range compressed motion dictionary on grid $\Theta$ using the 300 discriminant 
    vectors.
    \item Project all measured navigator signals into the discriminant subspace 
    (Eq.~\ref{eq:project}).
    \item Estimate motion parameters via dictionary matching at each contrast 
    (Eq.~\ref{eq:dot}).
    \item Refine estimates via continuous optimization for 100-150 epochs with temporal regularization 
    (Eq.~\ref{eq:refine}).
\end{enumerate}
The output is a rigid motion trajectory $\{\boldsymbol{\theta}_{(g,n)}\}$ of size 
$N_{\text{groups}} \times N_{\text{navi}} \times 6$.

The gEVD compression was implemented using the PyTorch Linear Algebra package~\cite{torchlinalg2021}. Motion-dictionary generation and dictionary matching were implemented in CuPy~\cite{cupy}. The regularization parameter in Eq.~\ref{eq:dot} was set to $\lambda_{DM} = 1\times10^{-2}$ and considered only the 30 neighboring maxima around the global dot-product maximum.

Continuous motion refinement was implemented in PyTorch~\cite{paszke2019arxiv} using automatic differentiation~\cite{linnainmaa1970representation}. Gradients from all acquisition groups and contrasts were accumulated and jointly back-propagated with the motion regularization term in Eq.~\ref{eq:refine}. The Adam optimizer was used with learning rates in the range $10^{-2}$--$10^{-1}$, and optimization was run for 20--60 epochs until motion updates got smaller than $10^{-6}$.

\begin{figure*}
\centerline{\includegraphics[width=0.9\textwidth]{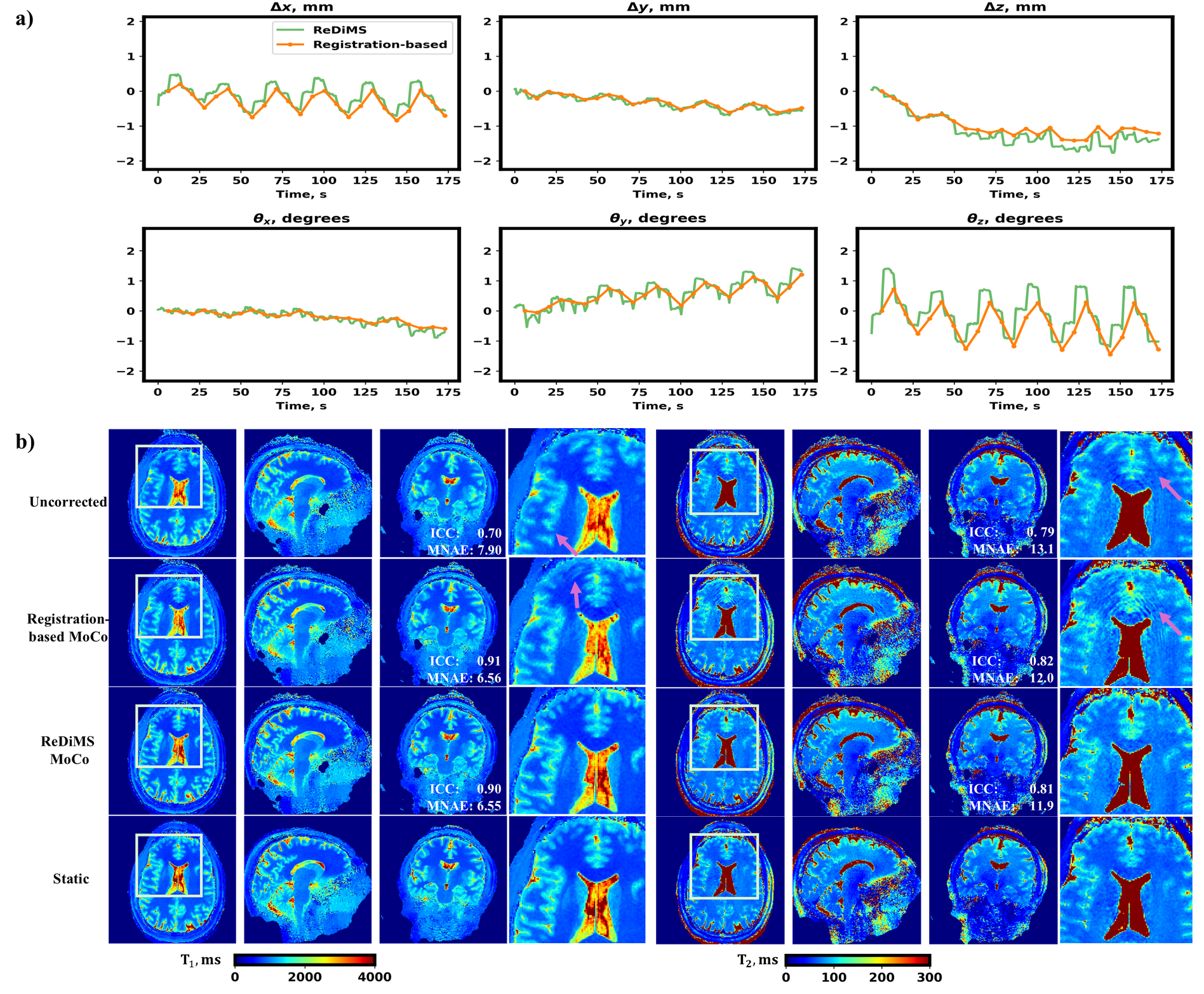}}
\caption{Motion estimation and correction results in case of small, frequent movements (mild motion) using image-registration-based and ReDiMS approaches. a) Estimated six rigid-body motion parameters illustrate that the temporal resolution of the image navigators is insufficient to capture rapid motion. b) Motion-corrected $T_1$ and $T_2$ tissue parameter maps, shown alongside uncorrected and static references. Registration-based motion correction reduces blurring compared to uncorrected maps, however, the registration-corrected T2 map contains more ringing artifacts. ReDiMS-based motion correction improves spatial resolution and reduces ringing artifacts, yielding maps that more closely resemble the static reference. \label{fig:small_motion}}
\end{figure*}

\begin{figure*}
\centerline{\includegraphics[width=0.9\textwidth]{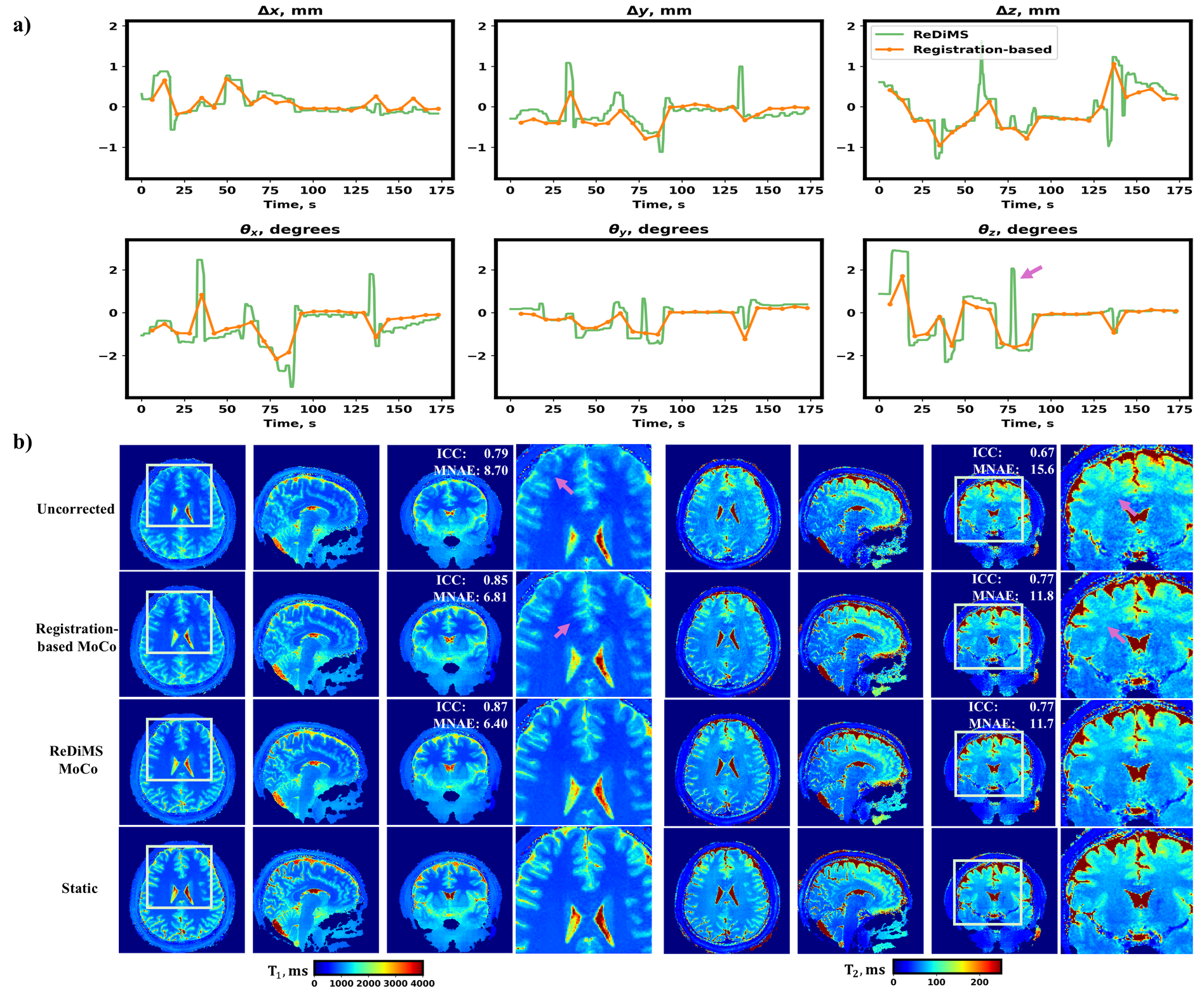}}
\caption{Motion-correction results for medium-severity, non-periodic motion using image registration-based and ReDiMS approaches. a) Estimated six rigid-body motion parameters show that image navigators capture most motion but may miss sharp transitions between acquisition groups. b) Motion-corrected $T_1$ and $T_2$ maps are shown alongside uncorrected and static references. Registration-based correction reduces artifacts, while navigator-based correction further improves spatial resolution, producing maps that more closely resemble the static reference. \label{fig:med_motion}}
\end{figure*}

\subsubsection{Simulations}
Simulation experiments were performed on static in vivo data from three subjects.

To evaluate ReDiMS design choices, navigator signal compressibility was analyzed using 
two motion dictionaries spanning 15\,000 states at grid step sizes of 0.5 and 
2.5\,mm/$^\circ$, measuring percent explained variance as a function of subspace rank 
for both SVD and gEVD bases. gEVD was further compared against SVD to characterize 
the effect of the basis transformation on navigator signal structure. Generalization 
of $U$ learned from a small motion grid $\{-0.5, 0, 0.5\}$ per DOF ($3^6 = 729$ 
states) was evaluated on navigator signals spanning 2--20\,mm/$^\circ$ using 
re-projection error and motion estimation accuracy.

To evaluate motion estimation performance, 500 random motion instances within 
$\pm12.5$\,mm/$^\circ$ were simulated per subject with Gaussian noise applied using 
the measured cross-channel covariance $\Sigma$. Five approaches were compared in 
motion estimation accuracy and compute time: optimization from zero initialization 
(60 and 300 epochs), dictionary matching only, and ReDiMS with 60 and 200 epoch refinement steps. No motion trajectory regularization was applied.

\subsubsection{In vivo}

Q-scout contrast consistency was evaluated across five test cases by comparing 
subspace coefficient maps reconstructed from the second dummy group data against 
those generated from the dummy preparation group data.

Discriminant signal projection effectiveness for in vivo motion estimation was evaluated using data acquired under instructed slow, step-wise motion across three subjects, comparing: (i) image-navigator registration-based estimation, (ii) raw navigator dictionary matching, (iii) SVD-compressed dictionary matching, (iv) gEVD-compressed dictionary matching, and (v) ReDiMS.

Full ReDiMS pipeline performance was evaluated across 35 motion-corrupted acquisitions from 10 subjects and compared with image-navigator registration-based estimation. Motion correction was assessed using NRMSE and SSIM on contrast-subspace coefficient maps. Evaluation was restricted to voxels with signal intensity above 10\% of the 98th percentile. 
Agreement between $T_1$ and $T_2$ maps measured under static and motion conditions was assessed using the Intraclass Correlation Coefficient (ICC)~\cite{Shrout1979} and mean normalized absolute error (MNAE, in percents) within white and gray matter regions segmented using~\cite{fischl2002whole,fischl2004sequence}. Agreement between registration-based and navigator-based motion trajectories was assessed using correlation analysis and mean absolute differences.

\section{Results}

\subsection{Simulations}
Simulations for ReDiMS design choices showed that gEVD required higher rank than SVD 
to capture 99.9\% of variance ($r = 81$ vs.\ $r = 13$; Figure S2a). 
Compressibility decreased with motion range (Figure S2c): rank increased 
from $r = 81$ for $\theta \in [-1,1]$\,mm/$^\circ$ to $r = 117$ for 
$\theta \in [-5,5]$\,mm/$^\circ$, motivating the selection of $r = 300$ to cover 
the full expected motion range $[-20,20]$\,mm/$^\circ$. Discriminant bases learned from a small motion grid 
$\{-0.5, 0, 0.5\}$ per DOF generalized well to navigator signals spanning 
$\pm(2$--$20)$\,mm/$^\circ$, with mean relative re-projection error below 13\% and 
motion estimation MAE below 5\% of maximum motion (Figure S2d), 
validating the use of a small dictionary for basis construction.

Motion estimation performance is summarized in Figure~\ref{fig:sims}. Across 500 simulated motion states, ReDiMS with 60-epoch refinement achieved MAE of $\sim$0.2\,mm/$^\circ$ in 7.0\,min, and extended 200-epoch refinement further reduced MAE to 0.02\,mm/$^\circ$ in 14.6\,min. Dictionary matching alone was faster but less accurate, with MAE of 2.4\,mm and 1.9$^\circ$ in 3.5\,min. 300-epoch optimization reached MAE of 0.52\,mm and 0.57$^\circ$ but required 16.7\,min. Zero-initialized 60-epoch optimization failed to converge within 3.3\,min, yielding MAE of 5.73\,mm and 4.35$^\circ$.

\begin{figure*}
\centerline{\includegraphics[width=0.9\textwidth]{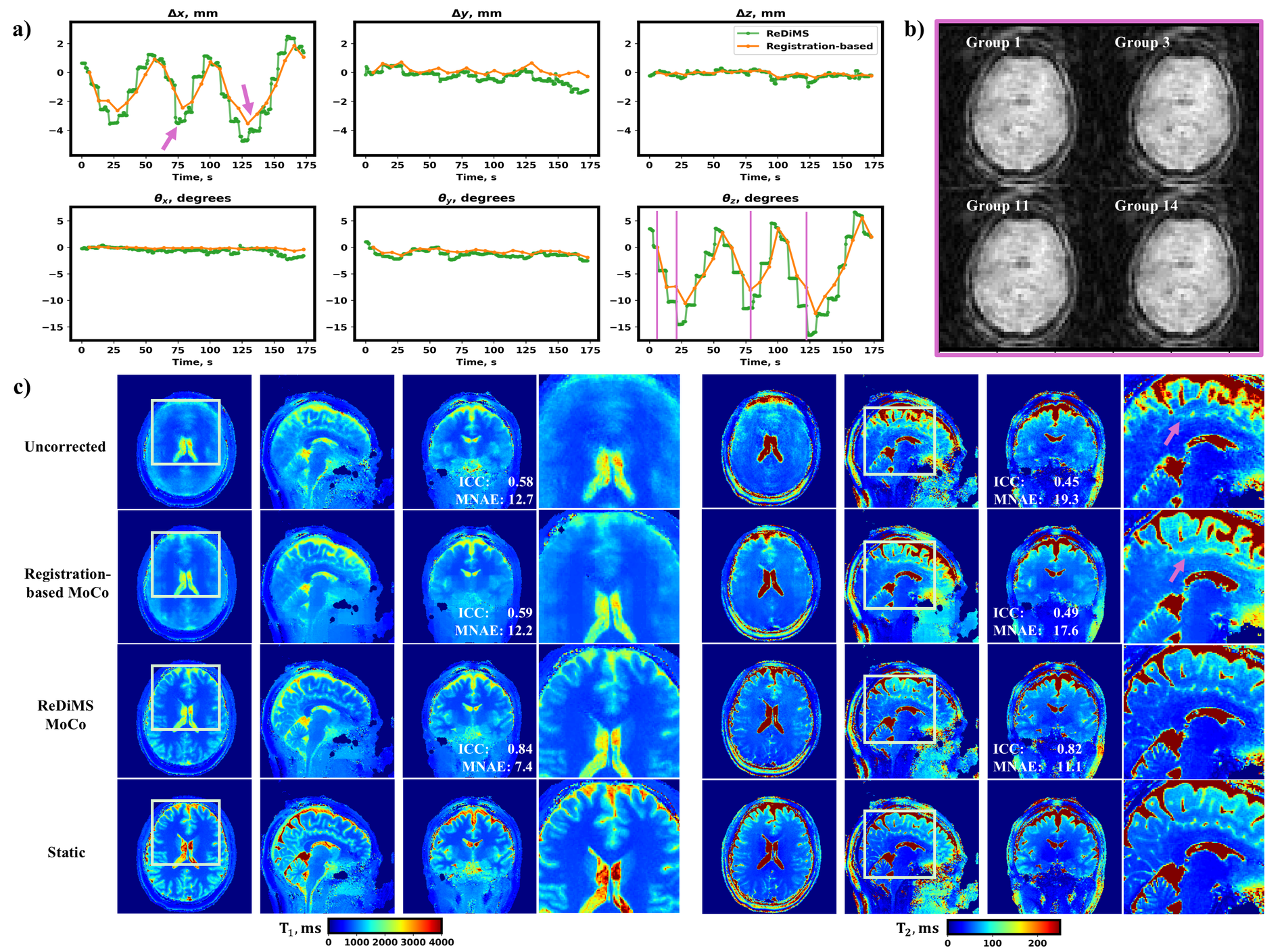}}
\caption{Motion-correction results for a severe motion case comparing image registration-based and ReDiMS approaches. a) Motion trajectories estimated by both methods show deviations at several timepoints (pink arrows). b) Representative image navigators (timepoints indicated by purple lines in a)) exhibit blurring reflecting potential motion during the 500\,ms navigator acquisition. c) $T_1$ and $T_2$ maps corrected by each method, alongside static and uncorrected references. Registration-based correction shows little improvement, whereas ReDiMS substantially recovers image quality, with residual blurring and CSF signal deviations remaining.\label{fig:large_motion}}
\end{figure*}

\subsection{In Vivo Tests}

Second dummy reconstruction and generated Q-scout subspace coefficient maps agreed within 5--8\%, with differences primarily localized to CSF and skull regions (Figure~\ref{fig:q_scout}).

To assess the impact of discriminant compression on dictionary matching-based motion estimation, dictionary matching estimates using raw, SVD-compressed, and gEVD-compressed signals were compared against registration-based reference trajectories (Figure~\ref{fig:compress}b). Raw signals showed substantial deviations and instability, SVD compression yielded similar performance, while gEVD compression markedly reduced deviations from the reference.

To validate the contribution of each ReDiMS stage, motion estimation and correction 
quality were assessed for raw dictionary matching, gEVD-compressed dictionary 
matching, and full ReDiMS (gEVD compression with optimization refinement), relative 
to a registration-based reference (Figure~\ref{fig:ablation}). Time-resolved 
reconstructions at TR index 125 show progressive artifact reduction across stages: 
raw dictionary matching already reduces motion artifacts relative to the uncorrected and registration-based MoCo
image, gEVD compression further reduces estimation bias, and optimization refinement 
corrects discretization errors, yielding image quality closest to the static 
reference. These trends are reflected by the motion trajectories in 
Figure~\ref{fig:ablation}b, and were consistent across additional cases in 
Figure S3.

Motion correction performance across different motion severities is shown in 
Figures~\ref{fig:small_motion}--\ref{fig:large_motion}. To enable comparison with 
the cohort-level statistics that follow, NRMSE is reported in-text on the 
reconstructed subspace coefficient maps relative to the uncorrected reconstruction. For small-amplitude periodic motion (Figure~\ref{fig:small_motion}), 
registration-based estimates failed to capture the motion dynamics, while ReDiMS 
recovered the trajectories and improved map sharpness, reducing NRMSE by 2.5\% 
versus 0.1\% for registration-based correction. For medium-strength aperiodic 
motion (Figure~\ref{fig:med_motion}), both methods reduced artifacts, with ReDiMS 
achieving a greater NRMSE reduction (8.2\% vs.\ 3.1\%). For strong motion 
(Figure~\ref{fig:large_motion}), registration-based correction showed limited 
improvement, whereas ReDiMS produced stable trajectories and substantially reduced 
NRMSE (14.0\% vs.\ 2.5\%). ICC and MNAE values in the figures further confirm the ReDiMS improvement 
over registration-based correction on the resulting $T_1$ and $T_2$ maps in 
the medium- and strong-motion cases, while remaining comparable in the 
small-motion case.

Across 35 motion-corrupted acquisitions, ReDiMS consistently outperformed registration-based 
correction, achieving mean NRMSE reductions of 7.1\% versus 3.9\% and SSIM 
increases of 0.085 versus 0.051, both statistically significant ($p = 1.9 \times 
10^{-8}$ and $p = 1.4 \times 10^{-7}$, respectively). Improvements were more 
pronounced in medium-to-severe cases, with mean NRMSE reductions of 8.0\% and SSIM 
increases of 0.103, compared to 6.1\% and 0.069 for mild motion. No deterioration in image quality was observed with ReDiMS, though 4 cases showed 
no measurable improvement similar to Figure S4. In contrast, several cases exhibited ringing artifacts 
following registration-based correction (Figure~\ref{fig:small_motion}).

Agreement between image-navigator registration and ReDiMS was assessed at overlapping 
timepoints. Estimated trajectories showed strong agreement, with mean correlation 
coefficients of 0.86 for translations and 0.89 for rotations, and mean absolute 
differences of 0.15\,mm and 0.23$^\circ$, respectively. Residual discrepancies are 
partially attributable to inaccurate motion estimation with image-navigator blurring under rapidly varying motion, as 
shown in Figure~\ref{fig:large_motion}.

\subsection{Compute Times}
Timing was evaluated for in vivo cases with $N_{\text{groups}} = 24$, 
$N_{\text{navi}} = 13$, 5 SVD-compressed coil channels, and three spiral navigators 
of total length $N_t = 4800$. The MRF image volume was $220\times220\times220$ with 
500 TRs, and the 4\,mm Q-scout $56\times56\times56$ with 5 subspace 
coefficients.

\begin{itemize}
    \item Dummy group subspace-LLR reconstruction (4\,mm-iso): $\sim$20\,s.
    \item Q-scout generation: 1--3\,s.
    \item Multi-contrast gEVD basis computation: $\sim N_{\text{navi}} \times 40$\,s 
    $\approx$ 8.7\,min.
    \item Motion dictionary generation (15\,000 states): $\sim N_{\text{coef}} \times 
    40$\,s $\approx$ 3\,min.
    \item Motion dictionary matching: 1--5\,s.
    \item Motion refinement via optimization: 1--4\,min (25--120 epochs, until motion 
    updates $< 10^{-6}$).
    \item Subspace-LLR reconstruction (1\,mm-iso): 5--7\,min.
\end{itemize}

\begin{figure*}
\centerline{\includegraphics[width=0.9\textwidth]{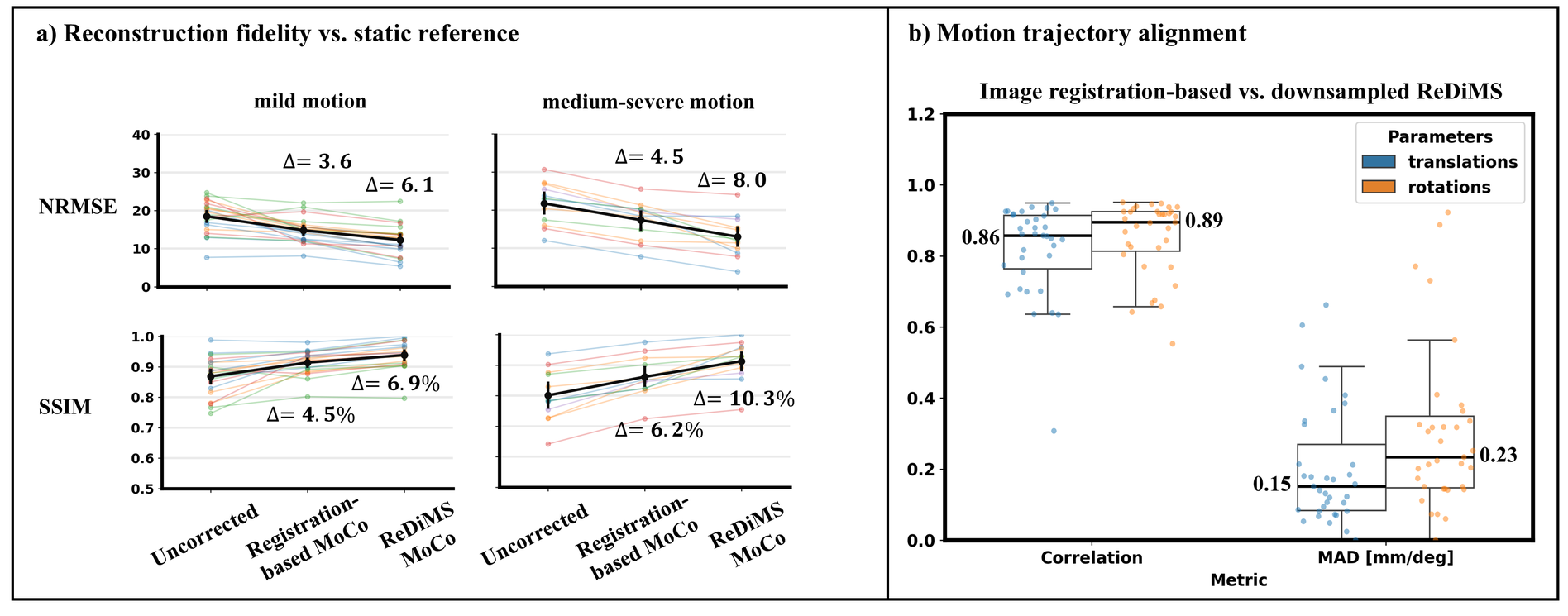}}
\caption{Motion estimation and correction performance across 10 volunteers and 35 motion cases. a) Reconstruction fidelity (NRMSE, SSIM) between static reference and uncorrected, registration-based, and ReDiMS-corrected reconstructions, grouped by estimated trajectory range (mild vs medium-severe), with representative examples in Figures~\ref{fig:small_motion}-\ref{fig:med_motion}. ReDiMS outperforms registration-based correction, reducing NRMSE by up to 6.1\% vs.\ 3.6\% (mild) and 8.0\% vs.\ 4.5\% (medium-severe), and increasing SSIM by up to 6.9\% vs.\ 4.5\% and 10.3\% vs.\ 6.2\%, respectively. b) Trajectory comparison between registration-based and ReDiMS estimates (latter is downsampled to matched timepoints). Agreement is high (correlation 0.86--0.89), with mean absolute differences of 0.15\,mm for translations and $\sim$0.23$^\circ$ for rotations. The long tail of the distribution likely reflects severe cases where image navigator corruption limits registration-based estimation accuracy. \label{fig:img_stats}}
\end{figure*}

\section{Discussion}
The registration-based correction used as reference operates on the same acquired sequence data, however, does not represent ground truth. Independent validation using external tracking systems, such as optical tracking or pilot-tone-based approaches~\cite{zaitsev2015optical,ludwig2021pilot,brackenier_pt,anand_bpt}, remains an important direction for future work. Image navigators in 3D SPI MRF could further be used in the proposed pipeline by initializing continuous optimization, serving as anchor points for regularization, constraining dictionary search ranges, or improving discriminant extraction. The latter would leverage reference motion estimates and corresponding in vivo navigator measurements at multiple poses to enrich the simulated-in vivo signal pairs used for discriminant construction.

While demonstrated in 3D SPI MRF, the Q-scout and compact navigator strategy extends to other acquisitions with evolving contrast, such as MPRAGE~\cite{Calakli2026_91e06143} or dynamic contrast imaging, and could guide spatio-temporal implicit neural representation reconstructions~\cite{inr_feng,nstm_motion} in a manner similar to~\cite{Li2025_NavigatorMotionResolved_MRF_INR}. The discriminant analysis framework is likewise applicable beyond motion estimation to problems with unavoidable forward-model mismatch, such as rapid calibration for field correction~\cite{Shah2026_4b46bc0e}.

The hybrid discrete-continuous optimization provides substantial speedup over purely continuous optimization, particularly when the motion optimization loss landscape is shallow and motion parameters are coupled. Analysis of our navigator setup in Figure S1 revealed moderate conditioning with condition numbers of 180--800 and noticeable coupling between motion parameters with Jacobian column correlations of 0.6--0.7. Such anisotropy leads to direction-dependent sensitivity and increased reliance on initialization with large motion extents, whereas dictionary matching performs a global search, avoiding dependence on local curvature.

Several limitations of the proposed approach should be noted. The framework assumes rigid motion and does not account for dynamic $\Delta B_0$ changes, eddy currents, non-rigid deformations, or motion-induced coil sensitivity variations~\cite{Brackenier2023QUEEN,serger2026motion,lin2025}. The Q-scout calibration requires approximately 5 seconds of motion-free data at the start of the scan, meaning early motion can degrade the reference used for navigation. Acquiring multiple scout segments throughout the scan, motion-correcting every partition and combining them into a single motion-free reference volume could mitigate this limitation. The full pipeline currently requires approximately 20 minutes and 10\,GB GPU memory for a 24-group acquisition, which could be reduced through forward model approximations, discriminant precomputation, or cross-subject scout generalization.

As with all retrospective correction approaches, large k-space gaps and signal 
evolution corruption from severe motion can fundamentally limit our correction quality 
even when motion estimates are accurate. Prospective correction~\cite{Maclaren_prospective,ZAITSEV201733,tisdall_prosp,Ulrich2023ServoNav} 
therefore represents a promising extension, where the Q-scout could enable forward 
model calibration and ReDiMS could be accelerated to provide real-time motion estimates within a 
feedback loop. Combining fast but potentially imperfect prospective updates with 
retrospective refinement of both motion estimates and image correction would further 
leverage the strengths of both approaches for improved artifact elimination\cite{maclaren_hybrid_moco}. Future 
work will target prospective correction with single-TR navigator trajectory design, reduced calibration time through accelerated Q-scout 
reconstruction and learning-based signal-to-motion forward mapping~\cite{drone,GONG2026103935}, 
aiming for few-second calibration and sub-second inference rates.

\section{Conclusions}
A sub-second temporal-resolution motion navigation framework was developed for 3D spiral-projection MRF using a quantitative scout and spiral navigator readouts with minimal sequence overhead. The proposed ReDiMS algorithm achieves computationally efficient and accurate motion estimation through discriminant projection, dictionary matching, and lightweight optimization refinement. In vivo evaluations demonstrated motion correction capability without introducing artifacts, and showed substantial improvements over image-navigator-based approaches, particularly in moderate-to-severe motion cases. These results suggest that high-temporal-resolution, contrast-aware motion navigation is a promising approach for improving the robustness of quantitative MRI under motion.

\section*{Acknowledgments}
The authors want to thank Mengze Gao, Julio Oscanoa, Cagan Alkan, Mark Nishimura for the helpful discussions driving this work.

\subsection*{Conflict of interest}

The authors declare no potential conflict of interests.

\bibliography{MRM-AMA}%

\end{document}